\documentstyle[11pt,newpasp,twoside,epsf]{article}
\def\edcomment#1{\iffalse\marginpar{\raggedright\sl#1\/}\else\relax\fi}
\reversemarginpar

\begin{document}
\title{On application of multi-colour photometry of $\delta$ Scuti stars}
\author{J. Daszy{\'n}ska-Daszkiewicz$^{1,2}$, W.A. Dziembowski$^{3,4}$, \\
A.A. Pamyatnykh$^{3,5,6}$ }
\affil{
$^1$Astronomical Institute of the Wroc{\l}aw University, Kopernika 11, 51-622 Wroc{\l}aw, Poland\\
$^2$Instituut voor Sterrenkunde, Katholieke Universiteit Leuven, Celestijnenlaan 200 B, B-3001 Leuven, Belgium\\
$^3$Copernicus Astronomical Center, Bartycka 18, 00-716 Warsaw, Poland\\
$^4$Warsaw University Observatory, Al. Ujazdowskie 4, Warsaw, Poland\\
$^5$Institute of Astronomy, Russian Academy of Sciences, Pyatnitskaya Str. 48, 109017 Moscow, Russia\\
$^6$Institute of Astronomy, University of Vienna, T\"urkenschanzstr.17, A-1180 Vienna, Austria
}

\begin{abstract}
In $\delta$ Scuti star models the photometric amplitudes and phases
exhibit a strong dependence on convection, which enters through
the complex parameter $f$, that describes the bolometric flux variation.
We present a new method of extracting $\ell$ and $f$ from
multi-colour data and apply it to several $\delta$ Scuti stars.
The inferred values of $f$ are sufficiently accurate to yield
an useful constraint on models of stellar convection.
\end{abstract}

\section{Introduction}
Until now the main application of  multi-colour photometry of $\delta$ Scuti
stars has been the determination of the spherical harmonic degree, $\ell$,
of the observed modes (e.g. Balona \& Evers 1999, Garrido 2000).
The amplitudes and phases from measurements in various passbands do indeed
contain a signature of the $\ell$ value, but there is more information
to be extracted from these quantities. In principle we may calculate
their theoretical counterparts. For this we need appropriate stellar models
and a linear nonadiabatic code for calculating model oscillations.
From such calculations we derive the complex ratio of the local flux variation
to the radial displacement at the photosphere, $f$. In the case of $\delta$ Scuti
stars, however, there is a large uncertainty arising from lack of adequate theory
of stellar convection. The uncertainity is reflected in a strong sensitivity
of $f$ to the mixing-length prameter, $\alpha$, which translates to calculated
mode positions in the amplitude ratio vs. phase difference diagrams.
This strong sensitivity to the treatment of convection is not necessarily
bad news, if we can determine simultaneously $\ell$ and $f$. For this aim
we need data from at least three passbands. If we succeed, then the $f$-value
yields a useful constraint on convection.
In addition, if the identified mode is radial, the multi-passband
data may be used to refine stellar parameters. Finally, the use
of radial velocity measurements in our method improves significantly
determination of $\ell$ and $f$.

\section{Method for inferring $f$ parameter from observations}
The presented method is based on $\chi^2$ minimization assuming trial
values of $\ell$. We regard the $\ell$ degree and the associated
complex $f$ value as the solution if it corresponds to $\chi^2$ minimum,
which is much deeper than at other values of $\ell$. We write
the photometric complex amplitude in the form of the linear
observational equation for a number of passbands, $\lambda$,
$${\cal D}_{\ell}^{\lambda} ({\tilde\varepsilon} f) +{\cal
E}_{\ell}^{\lambda} {\tilde\varepsilon} = A^{\lambda},\eqno(1)$$
where
$${\tilde\varepsilon}\equiv \varepsilon Y^m_{\ell}(i,0),~~~~
{\cal D}_{\ell}^{\lambda}f\equiv b_{\ell}^{\lambda}
D_{1,\ell}^{\lambda},~~~~
{\cal E}_{\ell}^{\lambda}\equiv
b_{\ell}^{\lambda}(D_{2,\ell}+D_{3,\ell}^{\lambda}).$$
On the right-hand side we have measured amplitudes, $A^\lambda$,
expressed in the complex form. The quantities to be determined are
$({\tilde\varepsilon} f)$ and ${\tilde\varepsilon}$.

If we have data on spectral line variations, the set of equations (1)
may be supplemented with an expression for the first moment,
${\cal M}_1^{\lambda}$,
$${\rm i}\omega R \left( u_{\ell}^{\lambda} + \frac{GM
v_{\ell}^{\lambda} }{R^3\omega^2} \right) \tilde\varepsilon
={\cal M}_1^{\lambda}.\eqno(2)$$
For more details and generalization of the method to the case
of modes coupled by rotation see Daszy\'nska-Daszkiewicz et al. (2003).
\begin{figure}
\mbox{\epsfclipon\epsfxsize=\textwidth\epsfysize=0.5\textwidth\epsfbox{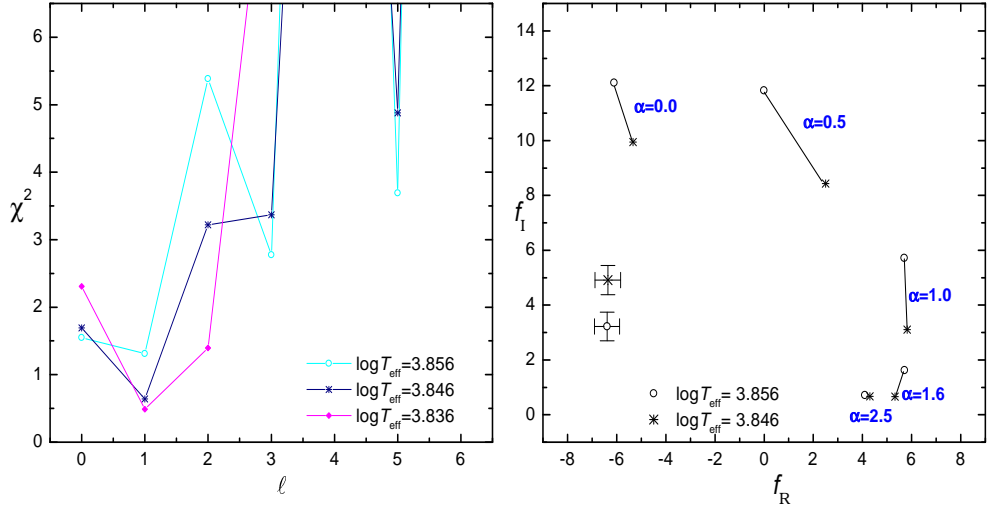}}
\caption{Left panel: $\chi^2$ as a function of $\ell$ for three models
of $\beta$ Cas. Right panel: comparision of empirical and theoretical
values of $f$ calculated for various values of $\alpha$ and two indicated
effective temperatures.}
\end{figure}
\section{Applications}
Here we rely on Kurucz (1998) models of stellar atmospheres
and Claret (2000) computations of limb-darkening coefficients.
For application of the method we used $uvby$ Str\"omgren photometry
in all cases.

{\bf $\beta$ Cas}: $\beta$ Cas is a near and bright $\delta$ Scuti star,
therefore we have rather precise parameters for it. Its estimated value
of mass is $1.95 M_{\odot}$. In the left panel of Fig.1 we show
$\chi^2$ as a function of $\ell$ obtained for three models. The minimum
at $\ell=1$ is the deepest one, particularly at lower $T_{\rm eff}$.
Because the star is a relatively rapid rotator ($v_{\rm rot} >70$ km/s),
we have considered the possibility that the mode is an  $\ell =0$ and 2
combination resulting from the rotatational coupling. We found, however,
that at no inclination is the $\chi2$ value as low as at a single $\ell=1$.
Thus, the latter identification is most likely.

In the right panel we show a comparison of the $f$ values inferred from
observations for $\beta$ Cas with the theoretical values calculated
with the five indicated values of the MLT parameter, $\alpha$.
The observed values of $f_R$ are closer to those calculated with
$\alpha=0$, however values of $f_I$ require rather higher values of $\alpha$.\\
\begin{figure}
\mbox{\epsfclipon\epsfxsize=\textwidth\epsfysize=0.49\textwidth\epsfbox{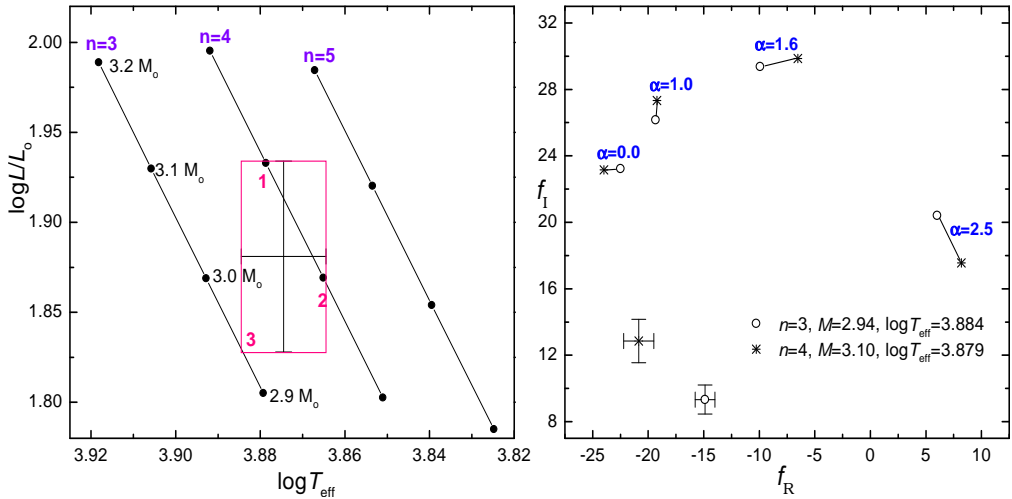}}
\caption{Left panel: the HR diagram with observational error box for 20 CVn.
Right panel: comparision of the empirical values of the $f$ parameter with
theoretical ones calculated for four values of $\alpha$.}
\end{figure}
%

{\bf 20 CVn}: 20 CVn is a $\delta$ Scuti variable regarded to be monoperiodic
with a metal abundance of [m/H]=0.5 $(Z\approx0.06)$.
For models on the edges of the error box at $Z=0.06$, the minimum
of the $\chi^2$ is clearly at $\ell=0$. Having such a mode identification
we can refine stellar parameters by fitting them to reproduce
the observed period.

In the HR diagram, shown in the left panel of Fig. 2, we plot the error box
and the lines of the constant period for radial orders $n=3,4,5$ at $Z=0.06$.
Only models along these lines are allowed. We cas see that we have two
possibilities: $n=3$ or $n=4$. The models yielding the
lowest $\chi^2$ are those marked with numbers 1 and 3, both equally probable.
In the case of $Z=0.04$ only $n=4$ is allowed, but the $\chi^2$ is much higher.
Thus from our method we have also some constraints on metallicity.
In the right panel we compare the empirical and theoretical values
of the $f$ parameter for models corresponding to the deepest $\chi^2$ minima.
The positions are qualitatively similar to those in $\beta$ Cas, but
considerably higher as a consequence of higher radial order.

{\bf 1 Mon}: 1 Mon has three almost equally distant frequencies. Here we
present results for the dominant mode. In the left panel of Fig. 5 we show
$\chi^2$ as a function of $\ell$ obtained from $uvby$ photometry,
for the models on the edges of error box.
\begin{figure}
\mbox{\epsfclipon\epsfxsize=\textwidth\epsfysize=0.46\textwidth\epsfbox{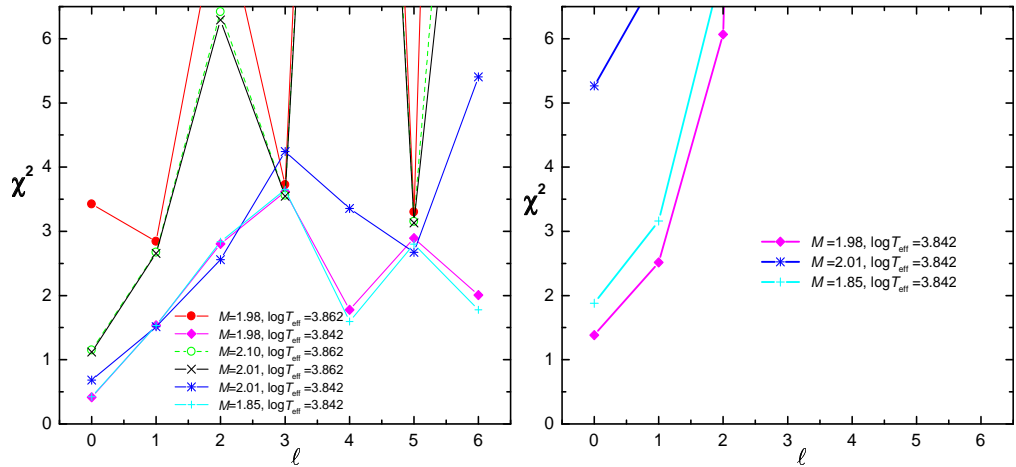}}
\caption{Left panel: $\chi^2$ as a function of $\ell$ from
photometry of 1 Mon. Right panel: the same after including
the radial velocity changes.}
\end{figure}
In the right panel we have the same, but after including data on
radial velocity changes. Discrimination is now much better.
Especially the higher values of $\ell$ and some sets of stellar
parameters are clearly excluded. The empirical values of the $f$ parameter
are now very close to those calculated with $\alpha=0.0$. We obtained
$f_{\rm obs}=(-4.9\pm 0.7,~11.3\pm0.8)$ and $f_{\rm calc}=(-2.4,~11.2)$.

\section{Conclusions}
We have argued that multi-colour photometry of $\delta$ Scuti stars
may give us useful information not only about the excited modes but
also about stars themselves. We believe that the most interesting is
the prospect of probing the efficiency of convective transport in outer
layers. We also showed that inferrence on mode degree and stellar
properties is much improved by supplementing photometric data
with radial velocity measurements.

\acknowledgements
The KBN grant No.5 P03D 012 20 is acknowledged.


\begin{references}
\reference Balona, L.A., Evers, E.A. 1999, \mnras, 302, 349
\reference Claret,  A. 2000, \aap,  363, 1081
\reference Daszy\'nska-Daszkiewicz, J., Dziembowski, W.A., Pamyatnykh, A.A. 2003, \aap, 407, 999
\reference Garrido, R., 2000, ASP Conf. Ser. 210, Delta Scuti and Related Stars,
eds. M. Breger\& M.H. Montgomery, (San Francisco ASP), 67
\reference Kurucz, R.L. 1998, http://cfaku5.harvard.edu
\end{references}
\end{document}